# Continuous and Orientation-preserving Correspondences via Functional Maps

JING REN, KAUST
ADRIEN POULENARD, LIX, École Polytechnique
PETER WONKA, KAUST
MAKS OVSJANIKOV, LIX, École Polytechnique

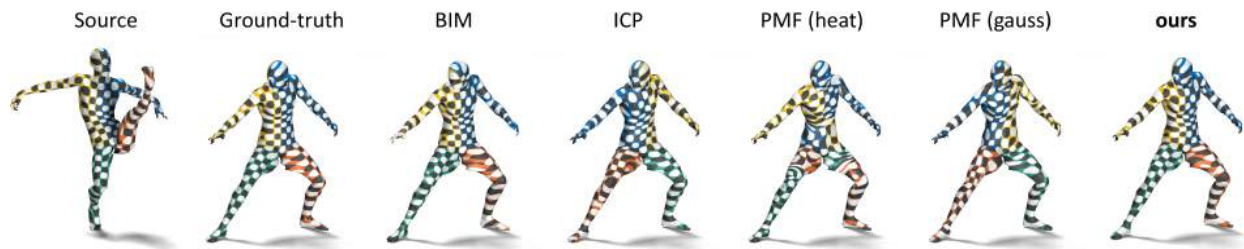

Fig. 1. Given a pair of shapes, our method produces a point-wise map that is orientation-preserving as well as approximately continuous and bijective. Here we show the maps produced by different methods via texture transfer: BIM [Kim et al. 2011] has a large distortion on the face and the left hand; functional maps with ICP [Ovsjanikov et al. 2012] and PMF with the Gauss kernel [Vestner et al. 2017b] give a map that is flipped left to right; for PMF with the heat kernel [Vestner et al. 2017a], the orientation in the torso region is reversed; The map produced by our method preserves the orientation consistently and has lower overall error when compared to the ground-truth.

We propose a method for efficiently computing orientation-preserving and approximately continuous correspondences between non-rigid shapes, using the functional maps framework. We first show how orientation preservation can be formulated directly in the functional (spectral) domain without using landmark or region correspondences and without relying on external symmetry information. This allows us to obtain functional maps that promote orientation preservation, even when using descriptors, that are invariant to orientation changes. We then show how higher quality, approximately continuous and bijective pointwise correspondences can be obtained from initial functional maps by introducing a novel refinement technique that aims to simultaneously improve the maps both in the spectral and spatial domains. This leads to a general pipeline for computing correspondences between shapes that results in high-quality maps, while admitting an efficient optimization scheme. We show through extensive evaluation that our approach improves upon state-of-the-art results on challenging isometric and non-isometric correspondence benchmarks according to both measures of continuity and coverage as well as producing semantically meaningful correspondences as measured by the distance to ground truth maps.

Authors' addresses: Jing Ren, KAUST, jing.ren@kaust.edu.sa; Adrien Poulenard, LIX, École Polytechnique, adrien.poulenard@inria.fr; Peter Wonka, KAUST, pwonka@gmail.com; Maks Ovsjanikov, LIX, École Polytechnique, maks@lix.polytechnique.fr.

**ACM Reference Format:**
Jing Ren, Adrien Poulenard, Peter Wonka, and Maks Ovsjanikov. 2018. Continuous and Orientation-preserving Correspondences via Functional Maps. *ACM Trans. Graph.* 37, 6, Article 248 (November 2018), 16 pages. https://doi.org/10.1145/3272127.3275040

## 1 INTRODUCTION

Computing correspondences or maps between shapes is one of the oldest problems in Computer Graphics and Geometry Processing with a wide range of applications from deformation transfer [Sumner and Popović 2004], statistical shape analysis [Bogo et al. 2014] to co-segmentation and exploration [Huang et al. 2014] among a myriad others. As a result a large number of approaches have been considered to tackle this problem in a wide variety of settings including rigid alignment [Besl and McKay 1992] and, more recently, in the more general case of computing correspondences between non-rigid shapes [Van Kaick et al. 2011].

A commonly used model for non-rigid shape matching is that of intrinsic isometries, which aims at finding correspondences that preserve geodesic distances between every pair of points as well as possible [Bronstein et al. 2006; Mémoli and Sapiro 2005]. Unfortunately, although this model has several attractive properties from the theoretical standpoint [Lipman and Funkhouser 2009], it also leads to very difficult optimization problems and corresponds, in full generality, to an NP-hard subclass of the quadratic assignment problem [Çela 2013; Solomon et al. 2016]. To alleviate this issue, recent methods have concentrated on either computing continuous correspondences by mapping the shapes into a common (parameterizing) domain, whose choice is primarily dictated by the shape topology, (e.g., [Aigerman and Lipman 2015, 2016; Lipman and Funkhouser 2009] among others), or, alternatively by relaxing the requirement for point-wise maps, by considering *generalized* or soft correspondences [Ovsjanikov et al. 2012; Solomon et al. 2012, 2016], which often leads to easier optimization problems.

The latter category has been particularly prominent, especially since methods built on the notion of soft mappings can take advantage of the recent advances in solving optimal transport problems, [Mandad et al. 2017; Solomon et al. 2015, 2016], or, alternatively, reduce to solving a simple least squares problem in the case of the functional maps framework [Kovnatsky et al. 2013; Ovsjanikov et al. 2012, 2017]. At the same time, while computing soft or functional maps can be done efficiently, extracting *continuous* or *bijective* point-wise maps is often challenging and error prone [Rodolà et al. 2015].

This problem is exacerbated even further by the presence of intrinsic symmetries, such as left to right symmetry, common to many shape classes, and which can lead to functional maps that represent a mixture between multiple possible solutions. Resolving this symmetry ambiguity is commonly done by assuming external information such as user-specified landmarks or training data in the context of learning [Boscaini et al. 2016], which further reduces the practical utility of the resulting pipeline.

In this paper, we address these challenges and propose an approach that uses the functional maps pipeline to produce a point-wise map that is orientation-preserving, approximately continuous and bijective, as illustrated, e.g., in Figure 1. Our extension has two main components. First, we introduce a new purely geometric *orientation-preservation energy term* for estimating functional maps, by exploiting intrinsic descriptor functions in a novel way. The second part is inspired by the recent iterative methods based on variance minimization [Mandad et al. 2017; Vestner et al. 2017b] whose goal is to produce transport plans that minimize local variance. We show how a similar idea can be used efficiently in an iterative scheme, which significantly extends the Iterative Closest Point refinement proposed in the original functional maps work [Ovsjanikov et al. 2012]. Namely, we show that by alternating between map optimization in the *spectral and spatial domains*, we can obtain a significant improvement in the overall map quality without relying on expensive, linear or quadratic, assignment problems, which allows us to handle more complex shapes efficiently. We demonstrate that our formulation can outperform state-of-the-art techniques on standard benchmarks containing near isometric and non-isometric shape pairs.

*Contributions.* To summarize, our contributions include:

(1) We propose a new term to promote orientation-preserving maps in the functional maps framework.
(2) We propose a new refinement scheme that improves the maps in both the spectral and spatial domains, while promoting continuity, coverage, bijectivity and while controlling for outliers that have a strong negative effect on prior methods.
(3) Through extensive experimental evaluation we demonstrate significant improvement over the state-of-the-art methods on standard benchmarks.

## 2 RELATED WORK

As shape matching is a very vast and well-studied area of Computer Graphics, below we review the methods most closely related to ours, concentrating on techniques aimed at producing continuous correspondences between non-rigid shapes, and refer the reader to recent surveys including [Biasotti et al. 2016; Tam et al. 2013; Van Kaick et al. 2011] for an in-depth treatment.

*Point-based methods.* Most early non-rigid shape matching methods concentrated on directly finding correspondences between points on the two shapes, while minimizing a prescribed energy, such as approximate preservation of geodesic distances, e.g., [Bronstein et al. 2006; Huang et al. 2008; Ovsjanikov et al. 2010; Sahillioğlu and Yemez 2010; Tevs et al. 2009] among many others. This model has appealing theoretical properties, since for example, it is known that a small number of landmarks is often sufficient to recover the full isometric map [Lipman and Funkhouser 2009; Ovsjanikov et al. 2010]. At the same time, these methods often lead to difficult combinatorial problems, and can easily fail when the intrinsic isometry assumption is violated. Moreover, although recent methods have tackled the computational complexity of quadratic assignment problems [Dym et al. 2017; Kezurer et al. 2015; Maron et al. 2016], they still remain computationally expensive, and often limited to tens or hundreds of points.

*Bijections through parameterization.* Another very fruitful line of work attempts to establish continuous maps between shapes by mapping them to a canonical domain, such as a sphere [Lipman and Funkhouser 2009] where continuous correspondences can be estimated explicitly from a small set of landmarks. Techniques that fall into this category have seen significant progress recently, both extending their applicability [Aigerman et al. 2017; Aigerman and Lipman 2016; Aigerman et al. 2015] to more general topological spaces and number of landmarks, as well as improving the computational complexity of the underlying methods. Nevertheless, they typically rely on a strict choice of landmarks and can potentially induce significant distortion in distant regions. A particularly successful extension of one of these techniques, called Blended Intrinsic Maps [Kim et al. 2011] aims to lift these assumptions and estimates a set of candidate maps, which are then blended together into a consistent, and often continuous correspondence. This method is fully automatic and can produce high quality maps for similar shapes. However, as we show below, it can also fail in aligning shape features, especially in challenging non-isometric cases.

*Functional Maps.* Another set of techniques that have been proposed for non-rigid shape matching is based on the notion of functional maps introduced in [Ovsjanikov et al. 2012] and extended significantly in follow-up works, including [Aflalo and Kimmel 2013; Ezuz and Ben-Chen 2017; Kovnatsky et al. 2013; Rodolà et al. 2017] among others (see [Ovsjanikov et al. 2017] for a recent overview). These methods assume as input a set of function correspondences, which can be derived from point-wise landmarks, or from automatically estimated region correspondences. They then estimate a functional map matrix that allows to transfer real-valued functions across the two shapes, which is then converted to a point-wise map. Although the first step reduces to the solution of a linear system of equations, this last step can be difficult and error prone [Ezuz and Ben-Chen 2017; Rodolà et al. 2015]. Furthermore, the produced map can fail to be either continuous or bijective, since the correspondence is usually computed in a particular direction (from a source to a target shape). As a result several recent extensions have tried to alleviate these limitations by e.g. modifying an input map using vector field flow [Corman et al. 2015], promoting bijectivity via adjoint regularization [Huang and Ovsjanikov 2017] or deblurring maps via projection onto an appropriate subspace [Ezuz and Ben-Chen 2017]. Nevertheless, the use of these techniques is still restricted to cases where the initial input is of sufficiently high quality and they fail under moderate non-isometries.

*Optimal Transport and Variance Minimization.* Another commonly used relaxation for matching problems is based on the formalism of optimal transport, which has recently been used for finding bijective and continous correspondences [Mandad et al. 2017; Solomon et al. 2016; Vestner et al. 2017b]. These techniques have benefited from the recent computational advances in solving large-scale transport problems, especially using the Sinkhorn method under entropic regularization [Cuturi 2013; Solomon et al. 2015]. Our method is inspired by two recent methods in this category [Mandad et al. 2017; Vestner et al. 2017b], which have been proposed to efficiently find bijective maps, while promoting continuity by iteratively solving optimal transport problems, whose cost promotes continuity with respect to the current solution. The iterative nature of these methods allows them to tackle large-scale problems efficiently and can result in approximately continuous and bijective maps even in challenging cases.

*Orientation Preserving Maps.* The most common approaches for computing orientation-preserving maps are based on using parameterizing domains, such as the sphere or the hyperbolic plane, where orientation changes can be detected directly, e.g. [Aigerman and Lipman 2016; Kim et al. 2011], among many others. Unfortunately, using such domains can also induce significant distortion when mapping across arbitrary shapes. Other approaches have used multi-scale solution analysis for detecting symmetric flips [Sahillioğlu and Yemez 2013], which can help avoid local orientation changes, but does not allow to control global orientation preservation. Finally, in the functional maps framework, symmetry disambiguation has been addressed either by transferring some known symmetry from one of the shapes [Ovsjanikov et al. 2013], using a vector field flow starting from a given orientation-preserving map [Corman et al. 2015], or by using learning with extrinsic descriptors [Boscaini et al. 2016]. Unfortunately all these approaches require some *a priori* information, which is not always available. Differently, we propose a purely geometric method for enforcing orientation preservation, which can be formulated directly in the functional (spectral) domain, using only an outward normal field on each shape.

## 3 PROBLEM SETUP AND MOTIVATION

The main goal of our work is to compute continuous, orientation-preserving and approximately bijective maps between a pair of shapes, having arbitrary discretizations, and without any external information, such as user-provided landmarks. We assume that all shapes are represented as manifold triangle meshes, possibly with boundary, but place no restriction on the shapes having the same number of vertices or consistent triangulations. In full generality, recovering accurate correspondences is a well-known difficult problem. In our work, we concentrate on shapes that undergo approximately isometric deformations, i.e. deformations that roughly preserve geodesic distances between pairs of points on each shape. Nevertheless, as we show below, in practice our method is capable of handling significant deviation from this model and can produce accurate correspondences even between non-isometric shapes.

To solve this problem we adopt the formalism and general strategy of the functional maps framework, introduced in [Ovsjanikov et al. 2012] and significantly extended in follow-up works, such as [Kovnatsky et al. 2013; Litany et al. 2017; Rodolà et al. 2017] among many others (see also [Ovsjanikov et al. 2017] for an overview). One of the advantages of this framework is its generality, as it can naturally handle different shape discretizations, and can incorporate a wide range of prior information such as descriptor or landmark constraints. For completeness, we first give a brief overview of the main steps involved in computing functional maps and then describe our key contributions to this framework.

*Basic Pipeline.* Given a pair of shapes, $S_1, S_2$ containing, respectively, $n_1$ and $n_2$ vertices, the basic pipeline for computing a map between them using functional maps, consists of the following main steps (see Chapter 2 in [Ovsjanikov et al. 2017]) :

(1) Compute a set of $k_1 << n_1$ and $k_2 << n_2$ basis functions on each shape, and store them as columns in matrices $\Phi_1, \Phi_2$.
(2) Compute a set of $q$ descriptor (also called probe) functions on the shapes, that are expected to be approximately preserved by the unknown map. Store their coefficients in the corresponding bases as columns of matrices $A_1, A_2$, of size $k_1 \times q$ and $k_2 \times q$ respectively.
(3) Compute the optimal *functional map* $C$ via:

$$C = \arg\min_X \|XA_1 - A_2\|_F^2 + \alpha \|\Delta_2 X - X\Delta_1\|_F^2, \quad (1)$$

where $\Delta_1$ and $\Delta_2$ are the Laplace-Beltrami operators of the two shapes expressed in the respective bases.
(4) Convert the functional map $C$ to a point-to-point one, possibly via an iterative approach, such as the Iterative Closest Point (ICP) in the spectral embedding.

The key step (3) in the pipeline above is aimed at finding a functional map that would approximately align the given descriptor functions, and also commute with the Laplace-Beltrami operators, which corresponds to finding isometric (preserving geodesic distances) correspondences. This step has been further extended e.g. using more powerful descriptor preservation constraints via commutativity [Nogneng and Ovsjanikov 2017] or using manifold optimization [Kovnatsky et al. 2016] among many others (see Chapter 3 in [Ovsjanikov et al. 2017]).

While powerful and efficient, this pipeline has a significant limitation, which reduces its utility in practice. Namely, although computing a functional map can be done efficiently in practice, obtaining a *continuous point-to-point map*, using this pipeline, has proven both challenging and error-prone.

The difficulty of recovering continuous correspondences with functional maps is fundamentally related to two challenges: first, the functional map constraints in step (3), such as descriptor preservation are typically intrinsic and do not disambiguate different possible solutions, which might be present e.g. due to the left-right symmetries of the shapes. Second, enforcing continuity in the point-wise map recovery in step (4) can quickly lead to difficult quadratic optimization problems. Furthermore, these two challenges are closely related, since e.g. the notion of an orientation preservation, which can be used to disambiguate possible symmetric correspondences is defined only for continuous maps.

In this paper we propose an approach to overcome these challenges, and to produce approximately continuous and orientation-preserving maps without any additional information, such as user-provided landmarks, knowledge of symmetries or training data for learning. For this, we first introduce a novel term designed to promote orientation-preserving maps, directly in step (3) of the initial functional map estimation pipeline. This term helps to reduce the ambiguity that might be present due to, e.g. left-right symmetries in the shapes, but is still not guaranteed to lead to continuous maps. To achieve this, we modify the pointwise map recovery, step (4), by introducing a bijective and continuous ICP (BCICP) approach, which unlike the basic ICP method, improves the maps *both in the spatial and in the spectral domains*, resulting in significantly higher quality functional and point-to-point correspondences.

## 4 PROPOSED METHOD

As mentioned above, our goal is to find orientation-preserving, approximately continuous and bijective point-wise maps. We first introduce an orientation-preserving energy term which directly fits into the functional map pipeline. We then introduce a refinement step, bijective and continuous ICP (BCICP) to convert the functional map to a point-wise one, where both bijectivity and continuity are strongly promoted, while refining the maps in both the spectral and spatial domains. We also propose a simple and efficient approach that places special emphasis on detection and removal of outlier correspondences, which commonly occur with existing methods and which can have strong impact on the visual and semantic quality. Below we describe each of these steps in our pipeline.

### 4.1 Orientation preservation

Many successful methods in shape matching, e.g. [Bronstein et al. 2006; Huang et al. 2008; Ovsjanikov et al. 2010; Sahillioğlu and Yemez 2010; Tevs et al. 2009] among others, are based on computing intrinsic isometries, which are maps that must approximately preserve intrinsic (geodesic) distances between pairs of points on each of the shapes. Although very powerful, this deformation model also has a significant limitation, since many shape classes naturally have *intrinsic symmetries*, which are non-identity self-maps that preserve geodesic distances. Such symmetries imply that multiple possible equally-likely correspondences might exist between pairs of shapes. Perhaps the most common source of ambiguities are the left-right symmetries present in both organic and man-made shapes (see e.g. Figure 2, left). Resolving such ambiguities has long been a challenge for purely intrinsic techniques, with several approaches being proposed based on learning [Boscaini et al. 2016] or using *a priori* knowledge of symmetries on some shapes [Ovsjanikov et al. 2013] among others. Below we propose a purely geometric technique to resolve these ambiguities, within the functional maps framework.

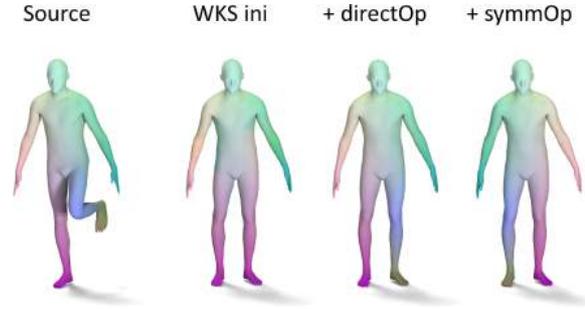

Fig. 2. Orientation preservation: for a pair of FAUST shapes, we used the Wave Kernel Signatures (WKS) as descriptors to initialize a functional map. Due to the symmetry ambiguity, the WKS initialization gives a map where both legs on the source are mapped to the left leg of the target. By adding the direct operator (+directOp) or the symmetric operator (+symmOp), we obtain the correct direct or symmetric maps from the WKS descriptors.

Our approach for deriving orientation-preserving constraints is based on the following classical observation: given two oriented surfaces $M, N$, a smooth map $T : M \to N$ is orientation preserving if and only if the map *differential* across tangent spaces $dT_p : \mathcal{T}_p M \to \mathcal{T}_{T(p)} N$ is orientation preserving [Guillemin and Pollack 2010]. In the case of embedded oriented surfaces, for any pair of vectors $v, w \in \mathcal{T}_p M$ in the tangent plane of $p \in M$, we can construct the triple product $\omega(v, w, \mathbf{n}(p)) = \langle v \times w, \mathbf{n}(p) \rangle$, where $\times$ is the standard cross-product and $\mathbf{n}(p)$ is the outward facing normal at $p$. Then, a smooth and bijective map is orientation preserving if and only if $\omega(v, w, \mathbf{n}(p))$ and $\omega(dT_p(v), dT_p(w), \mathbf{n}(T(p)))$, have the same sign (positive or negative), for all pairs of tangent vectors $v, w$ and for all points $p$. Moreover, in the case of isometric maps, this implies $\omega(v, w, \mathbf{n}(p)) = \omega(dT_p(v), dT_p(w), \mathbf{n}(T(p)))$.

One difficulty in translating this definition into an energy that can be optimized in practice is that it relies on the map differential $dT$ which is itself difficult to compute for maps between discrete surfaces that might not have the same triangulation. Moreover, enforcing the preservation of orientation directly might lead to non-convex energies that would be difficult to optimize.

Instead, we propose to approach this problem using the formalism of functional maps. Namely, given a functional map $F : \mathcal{F}(M) \to \mathcal{F}(N)$, across functional spaces of $M$ and $N$, so that $F(f) = g$, where $f : M \to \mathbb{R}$, and $g : N \to \mathbb{R}$, we consider the gradients, $\nabla f, \nabla g$ of $f$ and $g$. Furthermore, in the case of isometric maps the gradient commutes with the pull-back, which implies that if $F$ is a functional map associated with some pointwise isometry $T$, then $\nabla g = \nabla (F(f)) = dT_p(\nabla f(p))$, for all $p$. This means that in the context of isometries, orientation preservation can be written in the functional language as:

$$F(\Omega(\nabla f, \nabla h, \mathbf{n}_M)) = \Omega(\nabla F(f), \nabla F(h), \mathbf{n}_N) \; \forall \; f, h \in \mathcal{F}(M). \quad (2)$$

Here, $\mathbf{n}_M$ and $\mathbf{n}_N$ are outward normal fields on $M$ and $N$ respectively, $\Omega$ denotes the result of the triple product at each point, so that, e.g. $\Omega(\nabla f, \nabla h, \mathbf{n}_M)(p) = \omega(\nabla f(p), \nabla h(p), \mathbf{n}_M(p))$, and the equality must be understood as equality between functions.

In principle, Eq. (2), can directly be used as a constraint to enforce orientation preservation, when optimizing for a functional map $F$, by minimizing the difference between the two sides of the equation. This, however would lead to a difficult optimization problem since $F$ is present twice on the right side, meaning that terms involving products of functional maps would appear. Furthermore, the optimization would need to be taken over all pairs of functions, $f, h$, leading to a large number of energy terms.

An alternative approach is to observe that when the outward normal field $\mathbf{n}$ is fixed, $\Omega(\cdot, \cdot, \mathbf{n})$ is linear in each of its first two parameters, and $\Omega$ can be interpreted as a bilinear form mapping pairs of functions to real values through integration: $f, h \to \int \Omega(f, h, \mathbf{n}) d\mu$. This could allow us to define a single linear functional operator by duality with bilinear forms. Unfortunately, as shown in the following theorem, this bilinear form does not carry orientation information:

THEOREM 4.1. *For any closed surface:* $\int \Omega(f, h, \mathbf{n}) d\mu = 0 \; \forall h, f$

PROOF. See appendix □

To overcome these challenges, we observe that, the standard functional map pipeline is based on the presence of some descriptor (or probe function) correspondence constraints [Ovsjanikov et al. 2017]. These are pairs of functions $f_i, g_i$, typically obtained from some descriptors, such that we expect the unknown functional map to satisfy $F(f_i) = g_i$. Our observation is that if $f$ and $F(f)$ is replaced by $f_i$ and $g_i$ in Eq. (2) then this equation becomes linear in the functional map, and moreover the difference between the two sides can be interpreted as a linear functional operator, as follows: $F\bigl(\Omega(\nabla f_i, \nabla h, \mathbf{n}_M)\bigr) - \Omega(\nabla g_i, \nabla F(h), \mathbf{n}_N) = 0 \; \forall h \in \mathcal{F}(M)$ Finally, note that $\Omega(\nabla f_i, \cdot, \mathbf{n}_M)$ is a linear functional operator in its second parameter, which we denote as $\Omega_{f_i}(\cdot)$, with the normal field being implicit in the domain over which $f_i$ is defined. This leads to the following orientation preserving term:

$$\min_F \sum_i \left\| F \circ \Omega_{f_i} - \Omega_{g_i} \circ F \right\|^2 \qquad (3)$$

As observed in other works on functional maps, when linear operators are discretized as matrices in some basis, the composition $\circ$ becomes simply matrix multiplication. Using this term alongside other standard terms for functional maps, such as descriptor preservation and commutativity with the Laplace-Beltrami operators, allows us to recover maps that are both accurate and orientation preserving.

Finally, observe that if *orientation-reversing* maps are required, for example in the context of symmetry detection, this corresponds to reversing the orientation of the normal field, which in turn, by linearity, can be done simply by replacing the minus with a plus sign in Eq. (3). See Fig.3 for several examples of self-symmetric maps computed by reversing the orientation, starting with Wave Kernel Signature [Aubry et al. 2011a] descriptors, which are invariant to orientation changes.

*Initialization.* With this term at hand, we compute the initial functional map between a given pair of shapes $S_1$ and $S_2$, via:

$$C_{12} = \arg\min_X \alpha_1 \|XA_1 - A_2\|^2 + \alpha_2 \sum_i \|XD_{1i}^{\text{mult}} - D_{2i}^{\text{mult}}X\|^2 +$$
$$\alpha_3 \|\Delta_2 X - X\Delta_1\|^2 + \alpha_4 \sum_i \|XD_{1i}^{\text{orient}} - D_{2i}^{\text{orient}}X\|^2.$$

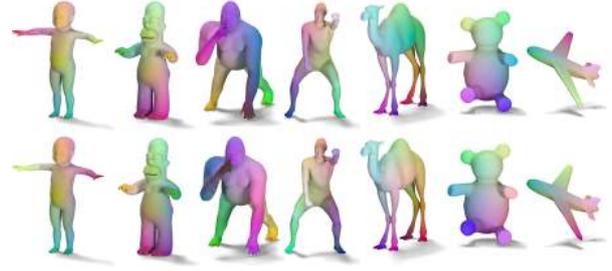

Fig. 3. Self-symmetric maps: our symmetric operator can be used to find the self-symmetric map of a shape, while using orientation-invariant WKS descriptors. The second row shows the self-symmetric maps of the corresponding shapes in the first row, via texture transfer.

Here, $D_i^{\text{mult}}$ are the multiplicative operators, introduced in [Nogneng and Ovsjanikov 2017], while $D_i^{\text{orient}}$ are the orientation-promoting operators $\Omega_{f_i}$ with respect to the $i^{\text{th}}$ descriptor function, as described in Eq. (3), but expressed as matrices in the given basis on the two shapes. Note that when using intrinsic descriptors, such as the Wave Kernel Signatures [Aubry et al. 2011b], only the last term promotes orientation preservation, without imposing assumptions on rigidity. We describe the exact choice of descriptors in Section 5.

### 4.2 Bijective and Continuous ICP (BCICP)

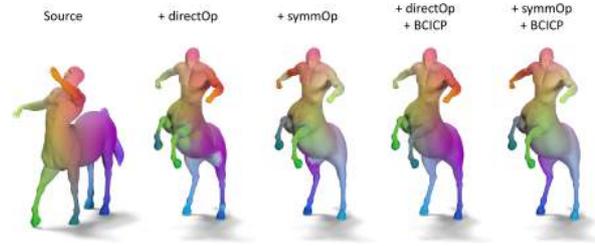

Fig. 4. Bijective and Continuous ICP: the orientation-preserving/reversing term is added as a soft constraint to regularize a map with preferred direction. However, it is not strong enough to guarantee that the produced map is free from symmetry ambiguity (such as the front leg region of the map produced with the direct operator or the arm region of the maps created with the symmetric operator). With the proposed BCICP refinement step, we can get a much better direct/symmetric map.

Although the orientation-preserving terms described above can help to reduce, e.g., left-right ambiguity present in many shape classes, it nevertheless does not necessarily lead to *continuous* or *bijective* maps. To achieve this, we introduce a novel approach based on iterative map refinement, by significantly extending the Iterative Closest Point (ICP) refinement proposed in the original functional maps pipeline [Ovsjanikov et al. 2012]. That method was based on iteratively recomputing the point-to-point map from its functional counterpart, simply by nearest neighbor search in the spectral domain, and updating the functional map by projecting onto the closest orthonormal matrix. Our main observation is that by updating the map *both in the spectral and spatial domains* we can obtain

**Algorithm 1** BCICP: Bijective and Continuous ICP

1: **Input** Initial functional maps $C_{12}^{(0)}, C_{21}^{(0)}$ and maxIter
2: **Initialization** $C_{12} \leftarrow C_{12}^{(0)}, C_{21} \leftarrow C_{21}^{(0)}$, iter $\leftarrow 1$
3: **while** iter $\leq$ maxIter **do**
4:     Compute vertex-to-vertex maps $\pi_{12}, \pi_{21}$    ▷ Algorithm 2
5:     Detect and fix the outliers in $\pi_{12}, \pi_{21}$    ▷ Algorithm 4
6:     Improve the coverage of $\pi_{12}, \pi_{21}$
7:     Improve the continuity of $\pi_{12}, \pi_{21}$    ▷ Algorithm 3
8:     $C_{12} \leftarrow \arg\min \|\Phi_2 C - \pi_{21}\Phi_1\|^2$
9:     $C_{21} \leftarrow \arg\min \|\Phi_1 C - \pi_{12}\Phi_2\|^2$
10: **return** bijective and continuous $\pi_{12}, \pi_{21}$, refined functional maps $C_{12}$ and $C_{21}$

**Algorithm 2** Bijective ICP for Functional Maps

1: **Input** Initial functional maps $C_{12}^{(0)}, C_{21}^{(0)}$ and maxIter
2: **Initialization** $C_{12} \leftarrow C_{12}^{(0)}, C_{21} \leftarrow C_{21}^{(0)}$, iter $\leftarrow 1$
3: **while** iter $\leq$ maxIter **do**
4:     $\pi_{21} \leftarrow \arg\min E'_1(\pi_{21} \mid C_{12})$
5:     $\pi_{12} \leftarrow \arg\min E'_2(\pi_{12} \mid C_{21})$
6:     $C_{11} \leftarrow \arg\min E'_3(C_{11} \mid \pi_{12}, \pi_{21})$
7:     $C_{11} \leftarrow \text{Proj}(C_{11})$
8:     $C_{22} \leftarrow \arg\min E'_4(C_{22} \mid \pi_{12}, \pi_{21})$
9:     $C_{22} \leftarrow \text{Proj}(C_{22})$
10:     $\pi_{12} \leftarrow \arg\min E'_3(\pi_{12} \mid C_{11}, \pi_{21})$
11:     $\pi_{21} \leftarrow \arg\min E'_4(\pi_{21} \mid C_{22}, \pi_{12})$
12:     $C_{12} \leftarrow \arg\min E'_1(C_{12} \mid \pi_{21})$
13:     $C_{12} \leftarrow C_{12}\text{Proj}(C_{21}C_{12})$
14:     $C_{21} \leftarrow \arg\min E'_2(C_{21} \mid \pi_{12})$
15:     $C_{21} \leftarrow C_{21}\text{Proj}(C_{12}C_{21})$
16:     iter $\leftarrow$ iter + 1
17: **return** $\pi_{12}, \pi_{21}, C_{12}, C_{21}$

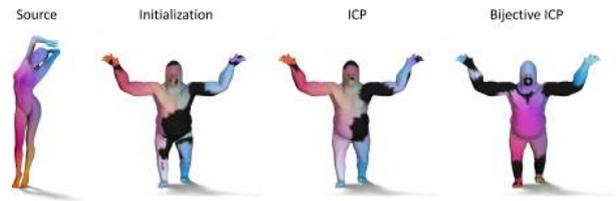

Fig. 5. Bijection improving accuracy. When the initial map contains large errors (e.g., the arms of the woman are mapped to the legs of the gorilla), it is hard for ICP to correct them. However, if we additionally consider the map from the other direction, we can improve the accuracy significantly. The regions colored black are outliers that are detected by the edge distortion. This example also illustrates that discontinuities and outliers are non-negligible problems in ICP.

significant improvement in the final overall map quality, and in particular, can promote continuity without sacrificing computational efficiency. Moreover, instead of processing a map in one direction, we consider two functional maps, in opposite directions between the two shapes, and process them jointly. This helps us to enforce invertibility, and thus the bijectivity of the resulting maps.

Our overall algorithm is summarized in Algorithm 1. We take as input two functional maps $C_{12}$ and $C_{21}$ between shapes $S_1$ and $S_2$, initialized using the approach described in the previous section, and alternate between computing two point-wise maps, refining them using the steps described in the following, and recomputing the induced functional maps.

In our paper, a point-wise map from $S_1$ to $S_2$ (with $n_1$ and $n_2$ vertices resp.) is represented in two ways: (1) as a vector $T_{12} \in \mathbb{R}^{n_1}$, where the $i$-th vertex on shape $S_1$ is mapped to $T_{12}(i)$-th vertex on shape $S_2$. We use this representation in the algorithm descriptions. (2) as a matrix $\pi_{12} \in \mathbb{R}^{n_1 \times n_2}$, where $\pi_{12}(i, T_{12}(i)) = 1, \forall i = 1, \cdots, n_1$, and the remaining entries equal to zero.

*4.2.1 Bijection.* Our first step is to modify the functional map refinement procedure to strongly promote bijectivity. For this we propose a simple iterative scheme that takes into account both the direct map $C_{12}$ and the one in the opposite direction $C_{21}$. A priori, these might not be inverses of each other, as, in the simplest case, they are initialized independently. Our goal, therefore is to rectify this loss of bijectivity while remaining close to the initial maps. See Fig. 5 for an illustrative example of how promoting bijectivity can improve the map between two shapes. For this, we construct the following energy:

$$E(C_{12}, C_{21}, \pi_{12}, \pi_{21}) = \lambda_1 E_1 + \lambda_2 E_2 + \lambda_3 E_3 + \lambda_4 E_4 \quad (4)$$

where

$$\begin{aligned} E_1 &= E_1(C_{12}, \pi_{21}) = \|\Phi_2 C_{12} - \pi_{21}\Phi_1\|_F^2 \\ E_2 &= E_2(C_{21}, \pi_{12}) = \|\Phi_1 C_{21} - \pi_{12}\Phi_2\|_F^2 \\ E_3 &= E_1(C_{11}, \pi_{12}, \pi_{21}) = \|\Phi_1 C_{11} - \pi_{12}\pi_{21}\Phi_1\|_F^2 \\ E_4 &= E_1(C_{22}, \pi_{12}, \pi_{21}) = \|\Phi_2 C_{22} - \pi_{21}\pi_{12}\Phi_2\|_F^2 \end{aligned} \quad (5)$$

Our goal is to solve the point-wise maps $\pi_{12}, \pi_{21}$ and refined functional maps $C_{12}, C_{21}$ from the following problem:

$$\min_{C_{12}, C_{21}, \pi_{12}, \pi_{21}, C_{11}, C_{22}} \lambda_1 E_1 + \lambda_2 E_2 + \lambda_3 E_3 + \lambda_4 E_4 \quad (6)$$

The first two terms $E_1$ and $E_2$ in (5) are the regular functional map energy defined on both directions. Bijectivity is promoted through the last two terms. Note that we introduce additional variables $C_{11}$ and $C_{22}$ describing self-maps under map composition. A natural possibility would be to try to enforce $C_{11}$ and $C_{22}$ to both equal identity. However, in practice, we have observed that especially when the initial maps are far from being inverses of each other, in an iterative scheme, it can be better to enforce a weaker constraint and require $C_{11}$ and $C_{22}$ to be *orthonormal* only. As observed in [Ovsjanikov et al. 2012; Rustamov et al. 2013] this constraint corresponds to requiring the underlying pointwise maps to be *area-preserving*, a property that the identity self-maps naturally satisfy.

The basic algorithm for bijective ICP refinement is summarized in Algorithm 2. Specifically, we do not optimize for all variables at the same time. For each iteration, we optimize each term in a sequence. Moreover, to speed it up, instead of solving a constrained problem, we decompose the problem into two steps: (1) first solve the unconstrained problem where we have a closed-form solution

**Algorithm 3** Continuity promotion: smooth a vertex-to-vertex map

1: $t^{(0)}_{\mathbf{x}_i} \leftarrow \mathbf{y}_{T(i)} - \mathbf{x}_i, \forall i = 1, \cdots, n_x$
   ▷ Initialize the displacement vector field
2: iter ← 1
3: **while** iter ≤ maxIter **do**
4:     **for** $i = 1, \cdots, n_x$ **do**
5:         $\mathcal{L}_i \leftarrow$ the largest connected component of $N^{\mathcal{Y}}_{T(N^{\mathcal{X}}_i)}$
   ▷ Update the search space
6:         $j \leftarrow \arg\min_{k \in \mathcal{L}_i} \left\| \left( \mathbf{x}_i + \frac{1}{|N^{\mathcal{X}}_i|} \sum_{l \in N^{\mathcal{X}}_i} t^{(\text{iter}-1)}_{\mathbf{x}_l} \right) - \mathbf{y}_k \right\|$
   ▷ Find the best candidate
7:         $t^{(\text{iter})}_{\mathbf{x}_i} \leftarrow \mathbf{y}_j - \mathbf{x}_i$   ▷ Update the correspondence field
8:         $T(i) \leftarrow j$   ▷ Update the vertex-to-vertex map
9: **return** $T$   ▷ the smoothed vertex-to-vertex map

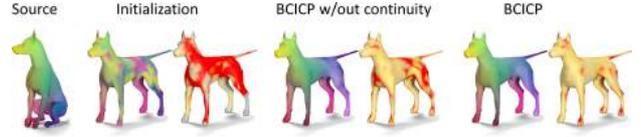

Fig. 6. Improvements due to continuity. We show maps for a single selected shape pair and visualize edge length distortions via a heat map (red indicates the strongest distortions). Given an initialization computed using functional maps, "BCICP w/out continuity" shows the result of our framework without the continuity improvement after a single iteration, while BCICP shows results with the continuity after a single iteration. Basic bijective ICP can smooth the noisy input map in the functional space, but the smoothness in the Euclidean space is not guaranteed, which is fixed by our approach.

(2) project the solution to the search space. For example, when we optimize for $C_{11}$ with given $\pi_{12}$ and $\pi_{21}$:

$$\min_{C_{11} \text{ is orthogonal}} E_3 = \left\| \Phi_1 C_{11} - \pi_{12}\pi_{21}\Phi_1 \right\|_F^2 \quad (7)$$

As discussed above, we can update $C_{11}$ in two steps as follows:

$$\begin{aligned} C_{11} &\leftarrow \Phi_1^\dagger \pi_{12}\pi_{21}\Phi_1 \\ C_{11} &\leftarrow \text{Proj}(C_{11}) \end{aligned} \quad (8)$$

where for a matrix $A$ with SVD decomposition $A = U\Sigma V^T$, the operator $\text{Proj}(\cdot)$ is defined as $\text{Proj}(A) = UV^T$. Similarly, we derive the update rules for the variables $C_{22}, C_{12}, C_{21}$. Finally, the steps in lines 13-15 of Algorithm 2 are designed to couple to computation of $C_{12}$ and $C_{21}$ by using a projection of their composition onto an orthonormal matrix. To derive these steps, we observe simply that if, e.g. $C_{12}$ satisfies $C_{12} = C_{12}\text{Proj}(C_{21}C_{12})$ it follows that $\text{Proj}(C_{21}C_{12}) = Id$. Moreover, if the composition $C_{21}C_{12}$ is orthonormal, this implies that the stationary points of this update rule must satisfy $C_{12}C_{21} = Id$. We discuss the role of this step in more detail in the Results section.

In addition, we can also incorporate the orientation-preserving term within the refinement step. Specifically, for each of the four terms, we can add an extra orientation term where the descriptors are constructed from the functional map. Take the first term $E_1$ as an example, $\Phi_1$ can be regarded as the descriptors (eigenfunctions) defined on shape $S_1$, while $\Phi_2 C_{12}$ are the corresponding descriptors (linear combinations of the eigenfunctions) defined on shape $S_2$. Therefore, similar to Eq. (2), we can construct the following energy with the orientation-preserving term:

$$E_1'(C_{12}, \pi_{21}) = E_1(C_{12}, \pi_{21}) + \alpha_1 \sum_i \left\| C_{12} \circ \Omega_{\Phi_1^{*i}} - \Omega_{\Phi_2 C_{12}^{*i}} \circ C_{12} \right\|_F^2 \quad (9)$$

where $A^{*i}$ denotes the $i$-th column of the matrix $A$. To simplify the computation, we can use $C_{12}$ from the previous iteration to construct the descriptors, then the energy is quadratic w.r.t the parameters $C_{12}$ and $\pi_{21}$. We can similarly add an extra orientation-preserving term to $E_i$ and denote the new energy as $E_i', \forall i = 2, 3, 4$.

*4.2.2 Continuity.* Unlike the basic ICP method that simply iterates between computing the point-wise and the induced functional maps, in addition to a new bijective term, we also introduce additional intermediate steps that promote continuity and the overall coverage. See Fig. 6 for an illustration how promoting continuity can influence the computed map. In our framework, after computing point-wise correspondences we modify them to rectify discontinuities, remove outliers and improve overall coverage.

In the first step, we use a similar correspondence smoothing technique to the Correspondence Estimation proposed in [Papazov and Burschka 2011]. The continuity is naturally preserved by mapping close-by vertices on the source shape to close-by vertices on the target shape which can be achieved by smoothing the correspondence vector field constructed from the map $T$.

Specifically, assume a point-wise map $T : \mathcal{X} \to \mathcal{Y}$ is given, where the shape $\mathcal{X}, \mathcal{Y}$ has the vertices' positions $\{\mathbf{x}_1, \cdots, \mathbf{x}_{n_x}\}$ and $\{\mathbf{y}_1, \cdots, \mathbf{y}_{n_y}\}$ stored in $X \in \mathbb{R}^{n_x \times 3}$ and $Y \in \mathbb{R}^{n_y \times 3}$ respectively. We let $N^{\mathcal{X}}$ denote the neighborhood structure of shape $\mathcal{X}$ so that $N^{\mathcal{X}}_i$ gives the list of neighbors of $\mathbf{x}_i$ vertex on shape $\mathcal{X}$. The notation for $\mathcal{Y}$ is identical.

The correspondence vector field $t \in \mathbb{R}^{n_x \times 3}$ is defined as the displacement vector between two corresponding vertices: $t_{\mathbf{x}_i} = \mathbf{y}_{T(i)} - \mathbf{x}_i$. Therefore, the average displacement at vertex $\mathbf{x}_i$ is computed as:

$$\bar{t}_{\mathbf{x}_i} = \frac{1}{|N^{\mathcal{X}}_i|} \sum_{k \in N^{\mathcal{X}}_i} t_{\mathbf{x}_k} \quad (10)$$

To smooth the correspondence at this vertex, we just need to find the nearest neighbor to $\mathbf{x}_i + \bar{t}_{\mathbf{x}_i}$ on the target shape. Instead of searching through the neighbors of $\mathbf{y}_{T(i)}$ as suggested in [Papazov and Burschka 2011], we restrict the search space to the largest connected component of $N^{\mathcal{Y}}_{T(N^{\mathcal{X}}_i)}$. Specifically, we first map the neighbors of $\mathbf{x}_i$, i.e., $N^{\mathcal{X}}_i$ to the target shape, and we get the set of vertices $T(N^{\mathcal{X}}_i)$. The neighbors of $T(N^{\mathcal{X}}_i)$ on shape $\mathcal{Y}$ are candidates for $\mathbf{x}_i$. To remove the possible outliers from the search space, we search through the largest connected component of $N^{\mathcal{Y}}_{T(N^{\mathcal{X}}_i)}$. Alg. 3 explains the steps to promote continuity in details.

*4.2.3 Outlier Regions.* In addition to the outlier removal step mentioned above, we also introduce an explicit procedure to detect

---

**Algorithm 4** Detect and fix the outliers of a map $T : \mathcal{X} \to \mathcal{Y}$

1: Get the edge list $\mathcal{E}$ from the adjacency matrix $A$ of the shape
2: **for** each edge $e = (e_1, e_2) \in \mathcal{E}$ **do**
3:     compute the edge distortion $r_e = \frac{\|\mathbf{y}_{T(e_1)} - \mathbf{y}_{T(e_2)}\|}{\|\mathbf{x}_{e_1} - \mathbf{x}_{e_2}\|}$
4:     **if** $r_e > \epsilon$ **then**
5:        $A(e_1, e_2), A(e_2, e_1) \leftarrow 0$    ▷ Break the connectivity
6: $\mathcal{V} \leftarrow$ the largest connected component of $A$
7: **for** $i = 1, \cdots, n_x$ **do**
8:     **if** $\mathbf{x}_i \notin \mathcal{V}$ **then**
9:        $T(i) \leftarrow T\big(\arg\min_{k \in \mathcal{V}} d(\mathbf{x}_i, \mathbf{x}_k)\big)$
                                 ▷ $d$ is the Geodesic or Euclidean distance
10: **return** $T$

---

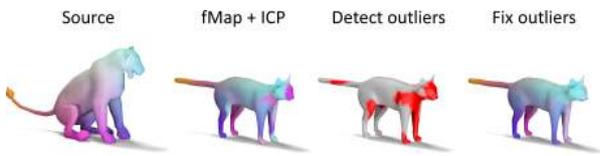

Fig. 7. Outlier Regions. Using functional maps can create outliers in the point to point correspondences. These outliers can be detected and fixed by our method to improve the mapping.

and remove outlier correspondence *regions* (See Fig. 7 for an example). These can be particularly problematic to existing methods, since they can affect the map globally making it difficult to recover a continuous correspondence.

To detect the outlier *regions*, we measure the distortion of every edge on the source shape, where the distortion is defined as the ratio of the length after and before the mapping. A large distortion suggests that at least one of the endpoints of this edge is an outlier. Therefore, such edges can help detect the boundary of the outlier regions. To detect the complete region of the outliers, we remove the edges with distortion above a certain (large) threshold from the adjacency matrix of the shape, and consider outliers to be all vertices that do not belong to the largest connected component according to this modified connectivity. Finally, we map each outlier vertex $v$ on the source shape to the same vertex that is mapped to by its nearest neighbor of $v$ in this largest connected component (see Algorithm 4). This helps to remove the outlier correspondences while the subsequent map coverage and continuity help to smooth and refine the resulting map.

*4.2.4 Coverage.* In addition to modifying the map to promote continuity and removing outlier regions from the correspondences, we also add a step to improve the *coverage* of each map, which is closely related to its bijectivity. The coverage of a vertex-to-vertex map is defined as the ratio between the area of the covered vertices and the total surface area of the target shape. A vertex on the target is called *covered* if there is at least one vertex on the source that is mapped to this vertex according to the vertex-to-vertex map.

We use the following simple heuristic to improve the coverage without hurting the local smoothness: Specifically, for each uncovered vertex $p$ on the target shape, we check if it has a neighbor that gets mapped to by multiple vertices from the source. We then find the neighbor with the largest preimage size, and pick a vertex from its preimage set as the preimage for $p$ (see Fig. 8). For this selection, we also consider the map in the opposite direction, to improve bijectivity. Please see the supplementary materials for the complete description of the heuristics that we use in this step.

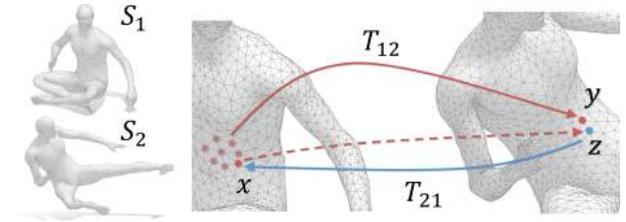

Fig. 8. Improving coverage: given point-wise maps $T_{12}$, $T_{21}$ between shapes $S_1$ and $S_2$, our goal is to use $T_{21}$ to improve the coverage of $T_{12}$ and vice versa. Thus, for any vertex $z$ on $S_2$ that is left *unmapped* by $T_{12}$, we find its neighbor $y$ with the largest pre-image size (here, all the red vertices are mapped to $y$ via $T_{12}$). Then if $T_{21}(z) = x$ we set $T_{12}(x) = z$.

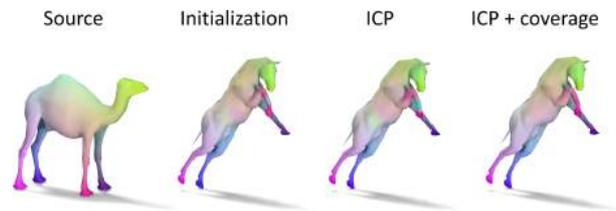

Fig. 9. Visualization of the coverage promotion step improving ICP. Note the incorrect mapping on the body and front fixed by our approach.

Fig. 9 illustrates the effect of including the coverage improvement step with the standard ICP functional map refinement. We observe, in particular, that step helps to better condition the linear system in functional map recovery, which allows to reconstruct more accurate vertex-to-vertex maps.

## 5 RESULTS

We implemented the proposed algorithms in MATLAB. In the following, we first describe the competing methods and their implementation. Then we describe the datasets and provide multiple quantitative and qualitative evaluations.

*Description of competing algorithms.* We compare our method, called *directOp + BCICP*, against multiple competing algorithms.

- directOp + BCICP: our complete framework as described in this paper, including the orientation-preservation term and the BCICP refinement steps.
- BIM: Blended Intrinsic Maps [Kim et al. 2011] using an executable provided by the original authors.
- (WKS/SEG) Initialization: The basic functional map framework, using the multiplicative operators proposed in [Nogneng and Ovsjanikov 2017] to compute the initial soft correspondences without any refinement, using as descriptors the wave kernel signatures (WKS) or the wave kernel maps built from region-based correspondences computed by an automatic algorithm [Kleiman and Ovsjanikov 2017].

- ICP: ICP to refine soft correspondences as described in the original functional maps framework [Ovsjanikov et al. 2012], and commonly used in follow-up works.
- PMF (gauss kernel): Product Manifold Filter [Vestner et al. 2017b] using the code of the original authors. They provide two versions of their code: 1) coarse-to-fine and 2) core algorithm initialized by a functional map. We use the version 2 (see below for an explanation).
- PMF (heat kernel): Product Manifold Filter using SHOT descriptors and Heat Kernel Signatures (HKS) to find the correspondences in the product space [Vestner et al. 2017a].
- CPD: Coherent point drift [Rodolà et al. 2015] using the implementation of the original authors.
- deblur: deblurring and denoising approach [Ezuz and Ben-Chen 2017] to post-process results to map vertices on the source shape to general surface locations on the target shape (within triangles), not only to other vertices.

In order to facilitate a reasonable comparison we omit hierarchical aspects of the different frameworks. One problem is that computation time can be prohibitive for conducting many experiments when high resolution data is used. In addition, the comparison is more difficult to interpret if factors relating to the hierarchical aspects of code influence the results. Therefore, we run the experiments on lower resolution data, so that hierarchical computation is not a requirement for any of the methods.

*Initialization.* For the methods listed above, all the approaches except BIM need given corresponding descriptors to compute a map. For PMF (heat kernel), the SHOT and HKS descriptors are used as suggested by the authors. For the other methods, we test two settings:

- **WKS**. Wave kernel signatures (WKS) are used as global descriptors for initialization.
- **SEG**. Region-based correspondences [Kleiman and Ovsjanikov 2017] are used to initialize functional maps. To control for the possible segmentation errors, we vary the parameter that controls the number of region correspondences, until the map, initialized from these region matches, has coverage above a certain threshold (25% in our experiments).

*Datasets.* We use multiple standard datasets to compare the different techniques: FAUST, TOSCA, and SHREC. To remove the bias present due to identical mesh connectivity within a dataset, we use LRVD algorithm [Yan et al. 2014] to remesh the datasets, while ensuring that each shape contains approximately 5k vertices. It is important to note that the **shapes were remeshed independently so that they do not share the same connectivity**, making our evaluation more difficult than in most prior works.

- FAUST: The FAUST dataset [Bogo et al. 2014] contains meshes of ten different humans in ten different poses each. The dataset can be split into two different types of shape pairs: isometric pairs stemming from the same person and non-isometric pairs stemming from two different persons.
- TOSCA: The TOSCA dataset [Bronstein et al. 2008] contains 80 meshes of humans and animals in 9 categories. It can naturally be split into two different types of shape pairs. Isometric pairs contain two shapes of the same class and *non-isometric pairs* are two shapes of a different class with dense manually verified correspondences.
- SHREC: We use the FourLegs category of the SHREC 2007 dataset [Giorgi et al. 2007].

*Measurement.* In our experiments, we measured the accuracy, continuity (smoothness), and bijectivity of the computed maps.

- **Accuracy**. In each dataset, the ground-truth direct and symmetric correspondence are given. To measure the accuracy of a computed map, we adopted the following measures:
  – **per-vertex error**: for each vertex we accept the ground-truth direct and symmetric correspondences and take the minimum as the error of this vertex.
  – **per-map error**: we compute the average per-vertex error to the direct map and the average per-vertex error to the symmetric map and take the minimum of these two metrics.
  – **direct error**: we compute the average per-vertex error to the direct ground-truth correspondences only.
- **Continuity and coverage**. To measure the map continuity, for each edge on the source shape, we compute the ratio between the geodesic distance of the two mapped endpoints on the target shape and the edge length. Note that a trivial map as mapping all the vertices to the same vertex on the target would have a low distortion, since nearby vertices vertices are mapped nearby. Therefore, for practical purposes, continuity has to be considered in conjunction with coverage, which is defined as the ratio between the area of the covered vertices and the total surface area of the target shape.
- **Bijectivity**. To measure bijectivity, we consider the maps from both sides, denoted as $T_{12}$ and $T_{21}$, and compute the average geodesic distance between the composed maps $T_{12} \circ T_{21}, T_{21} \circ T_{12}$ and the identity map.

### 5.1 Evaluation for automatic shape matching

In this sequence of tests, we compare the performance of our method in the context of automatic shape matching. We use functional maps with two different settings: (1) WKS and (2) SEG. We then add the orientation-preserving term and the BCICP refinement as our main method and denote it with (WKS/SEG + directOp + BCICP). We report the average geodesic error w.r.t the three error measures in Tables 1–4. Figures 10–13 show the corresponding cumulative curves of the percentage of matches that have error smaller than a threshold, when measured according to the three different error metrics. Note that when compared to the baseline methods (BIM and SEG + ICP) and with respect the direct error measure, our method achieves 17.5% improvement on 200 FAUST isometric dataset, 18.4% on 400 FAUST non-isometric dataset, 38.8% on 568 TOSCA isometric dataset, and 43.5% on 190 TOSCA non-isometric dataset. Moreover, even for the per-vertex error, our method achieves 17.8% improvement of accuracy on FAUST isometric pairs, 24.1% on FAUST non-isometric pairs, 31.4% on TOSCA isometric pairs, and 38.3% on non-isometric pairs. In summary, our algorithm (+ directOp + BCICP) significantly outperforms the other methods consistently across the FAUST and TOSCA dataset on isometric or non-isometric pairs.

Table 1. 200 FAUST Isometric pairs. We compare our algorithm with orientation-preservation and BCICP to competing methods: BIM, PMF with heat kernel and the ICP refinement. For both initializations, using WKS or SEGS, our algorithm outperforms the other methods.

| Methods | Ave geodesic error($\times 10^{-3}$) | | |
|---|---|---|---|
| | per vertex | per map | direct |
| BIM | 43.69 | 44.43 | 93.17 |
| PMF (heat kernel) | 57.89 | 61.66 | 62.06 |
| WKS + ICP | 42.52 | 120.65 | 150.84 |
| SEG + ICP | 27.52 | 30.09 | 30.09 |
| SEG + BCICP | 24.93 | 27.22 | 27.22 |
| **WKS + directOp + BCICP** | 25.94 | 39.28 | 44.16 |
| **SEG + directOp + BCICP** | **22.63** | **24.83** | **24.83** |

Table 2. 400 FAUST non-Isometric pairs.

| Methods | Ave geodesic error($\times 10^{-3}$) | | |
|---|---|---|---|
| | per vertex | per map | direct |
| BIM | 45.49 | 46.30 | 79.38 |
| PMF (heat kernel) | 55.92 | 59.95 | 60.54 |
| WKS + ICP | 65.59 | 166.37 | 210.11 |
| SEG + ICP | 30.94 | 45.68 | 45.68 |
| SEG + BCICP | 25.21 | 39.77 | 39.77 |
| **WKS + directOp + BCICP** | 29.33 | 46.90 | 51.31 |
| **SEG + directOp + BCICP** | **23.50** | **37.29** | **37.29** |

We note that running PMF with gauss kernel [Vestner et al. 2017b] on meshes with 5k vertices till convergence is computationally expensive (taking between 50 minutes and 7.5 hours depending on the shape pair and initialization). Therefore, here we only compare to PMF with heat kernel, and the results of PMF with gauss kernel in 10 iterations are shown in the supplementary materials. When comparing the two initializations, we can find that for FAUST, SEG is better than WKS, but in the case of TOSCA dataset, it is the opposite. The reason is that the region matching algorithm can fail for animal shapes since its parameters were tuned for human shapes. For the TOSCA dataset, there are segmentation errors like mapping the tail to the back leg, or mapping the head to the front leg. Therefore, for TOSCA dataset, the WKS is more stable than SEG. As illustrative visualizations, Figure 14 shows an example of one TOSCA isometric pair, where our method produces the direct and the symmetric map. Figure 15 gives another example of a non-isometric pair in the TOSCA dataset. The complete comparison (including the fMap initializations without any refinement, PMF with gauss kernel, and our method with orientation-reversing term) can be found in the supplementary materials.

### 5.2 Evaluation of BCICP w.r.t each component

From the results, it is easy to observe the usefulness of the orientation-preserving operators. Specifically, using WKS descriptors with direct/symmetric orientation-preserving/reversing operator, we can obtain a good direct/symmetric map. Also for SEG initialization, "+directOp + BCICP" performs consistently better than "+ BCICP," which verifies the usefulness of the orientation-preservation term.

Table 3. TOSCA Isometric: 568 pairs, including 132 pairs of Victoria, 120 pairs of Michael, 110 pairs of cats, 72 pairs of dogs, 52 pairs of horses, 40 pairs of David, 32 pairs of centaur and 6 pairs of wolfs.

| Methods | Ave geodesic error($\times 10^{-3}$) | | |
|---|---|---|---|
| | per vertex | per map | direct |
| BIM | 41.20 | 45.14 | 70.76 |
| PMF (heat kernel) | 49.03 | 71.44 | 71.82 |
| WKS + ICP | 44.90 | 128.45 | 194.01 |
| SEG + ICP | 42.93 | 61.36 | 73.30 |
| SEG + BCICP | 36.55 | 54.29 | 65.86 |
| **WKS + directOp + BCICP** | **28.25** | **38.56** | **43.34** |
| **SEG + directOp + BCICP** | 34.80 | 52.13 | 61.36 |

Table 4. 190 TOSCA non-Isometric pairs, including 120 pairs of gorilla and Victoria, and 70 pairs of gorilla and David.

| Methods | Ave geodesic error($\times 10^{-3}$) | | |
|---|---|---|---|
| | per vertex | per map | direct |
| BIM | 256.11 | 265.83 | 358.99 |
| PMF (heat kernel) | 190.21 | 249.14 | 287.29 |
| WKS + ICP | 314.82 | 406.67 | 449.75 |
| SEG + ICP | 145.92 | 178.14 | 214.05 |
| SEG + BCICP | 120.45 | 150.71 | 187.38 |
| **WKS + directOp + BCICP** | **90.03** | **112.68** | **120.96** |
| **SEG + directOp + BCICP** | 114.94 | 144.45 | 180.39 |

In addition, we evaluate the utility of the different components of the BCICP framework as discussed in Sec. 4 (see Table 5). We tested 50 pairs of TOSCA isometric pairs with initialization "WKS +directOp" (Ini). Then we run the regular ICP and the bijective ICP (bi-ICP), which is a simplified version of our BCICP approach, as discussed in Algorithm 2. The bijective ICP has a better performance than the regular ICP w.r.t the accuracy. Moreover, we also tested the settings where the coverage-improving step is removed (BCICP-cov.), and the smoothness-improving step is removed (BCICP-cont.), which includes the step of removing the outliers and improving the continuity. The results show that each component makes a contribution when compared to the regular ICP refinement. In addition, when they are combined together, we can achieve a much better result (last column of Table 5).

### 5.3 Improving coverage and smoothness using BCICP

A major advantage of our framework is to improve the coverage and smoothness of the resulting maps compared to the state of the art. On the one hand, coverage and smoothness work as a regularizer to reduce the error in the matching itself. On the other hand, coverage and smoothness are desirable attributes for a range of applications, including texture transfer and deformation transfer. In Fig. 16 we report results for the FAUST dataset and in Fig. 17 we report results for the TOSCA dataset. We can see that our approach drastically improves coverage as well as bijectivity on both datasets. In Fig. 18 we report a smoothness error metric evaluated on FAUST and TOSCA. All tests reveal that BCICP brings drastic improvements in terms of coverage and bijection compared to BIM and ICP. In terms of continuity, BCICP is comparable to BIM and ICP. However, the reason for the relatively good smoothness value of BIM or ICP is that they have low coverage with many vertices on the source shape mapped to the same vertex on the target shape. This gives zero edge distortion. For practical purposes, continuity has to be considered in conjunction with coverage.

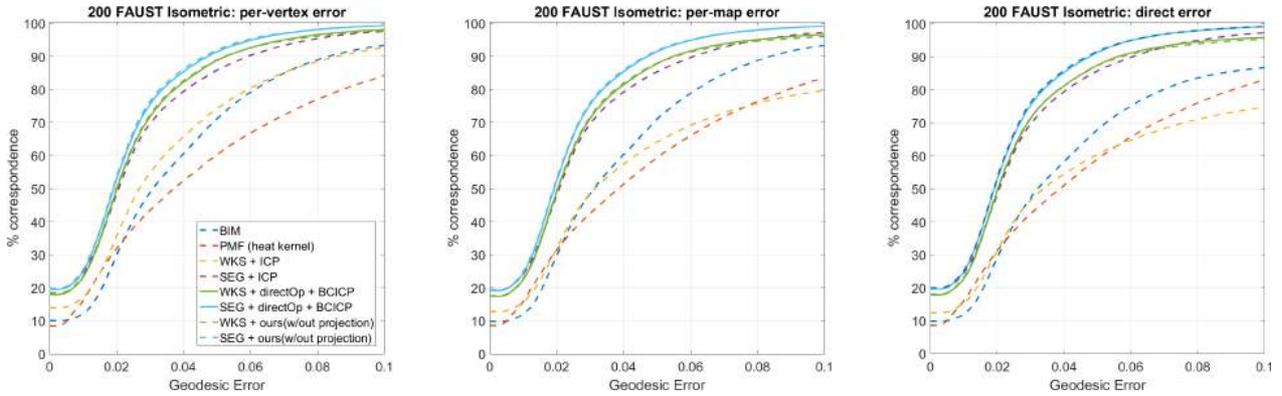

Fig. 10. Evaluation on 200 FAUST Isometric pairs with respect to three error measures.

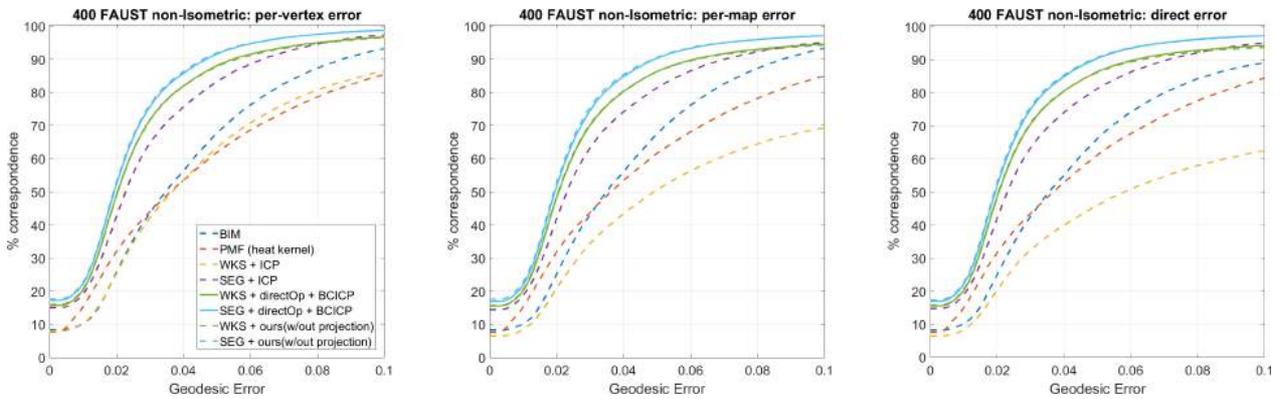

Fig. 11. Evaluation on 400 FAUST non-Isometric pairs with respect to three error measures.

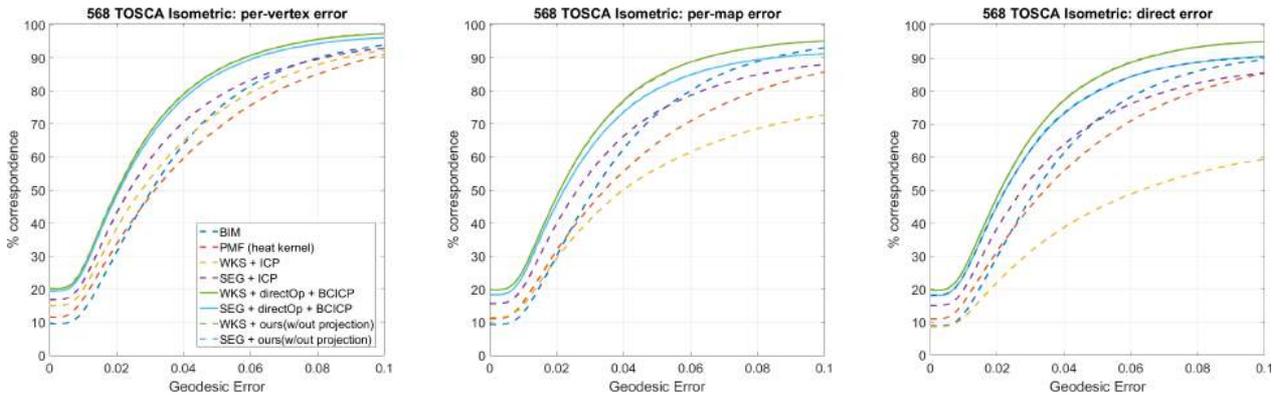

Fig. 12. Evaluation on 568 TOSCA Isometric pairs with respect to three error measures.

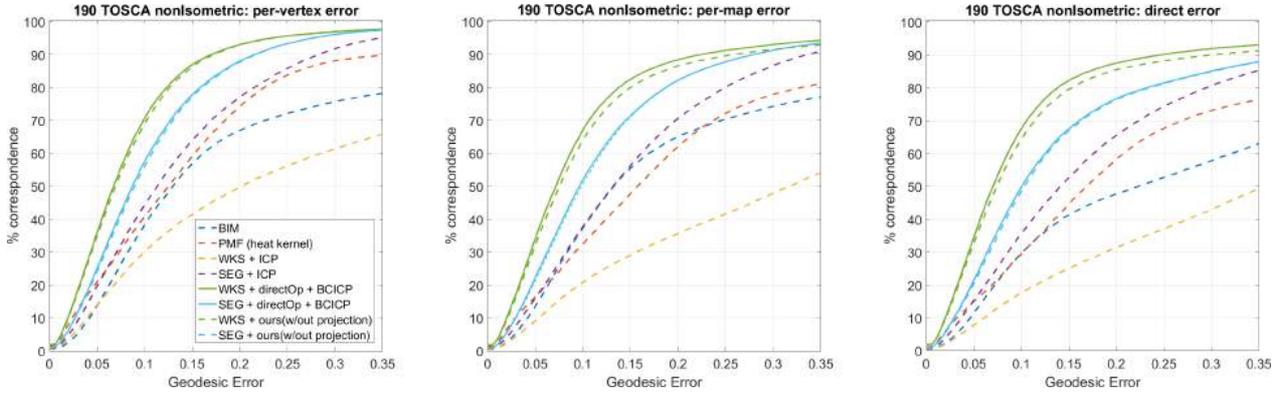

Fig. 13. Evaluation on 190 TOSCA non-Isometric pairs with respect to three error measures.

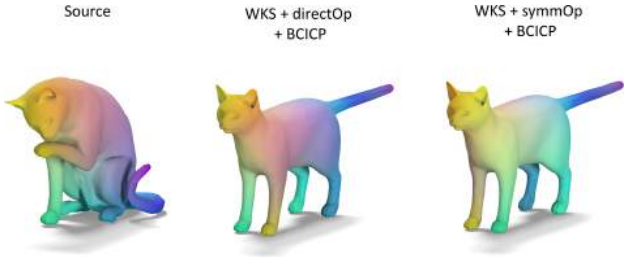

Fig. 14. An example isometric pair from the TOSCA dataset using the direct/symmetric orientation operator with BCICP refinement.

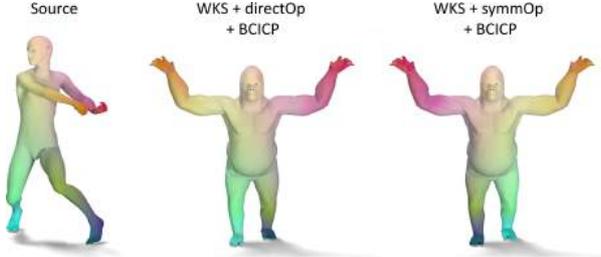

Fig. 15. An example non-isometric pair from the TOSCA dataset using the direct/symmetric orientation operator with BCICP refinement.

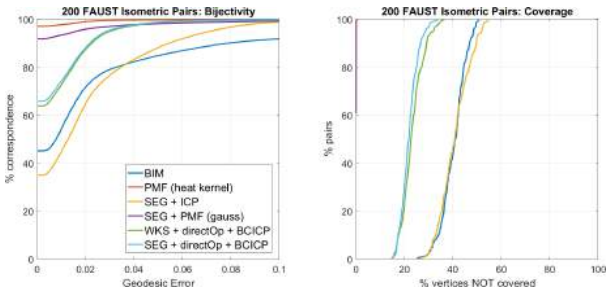

Fig. 16. Bijectivity (distance to identity) and coverage comparison on the FAUST dataset: 100 pairs of isometric shapes.

Table 5. Evaluation of each component of BCICP: we add bijectivity to ICP (bi-ICP) and removed coverage (BCICP - cov.) and continuity (BCICP - cont.) from the BCICP refinement. We evaluate each component with respect to different metrics. Note that ICP has the best continuity because it has a smaller coverage rate.

| Measure | Ini | ICP | bi-ICP | BCICP - cov. | BCICP - cont. | BCICP |
|---|---|---|---|---|---|---|
| Err - perVtx | 51.67 | 41.13 | 39.01 | 38.33 | 37.29 | **29.97** |
| Err - perMap | 67.52 | 52.43 | 50.04 | 45.09 | 47.59 | **34.87** |
| Err - direct | 67.86 | 52.43 | 50.04 | 45.08 | 47.59 | **34.87** |
| Coverage(%) | 40.05 | 54.40 | 55.75 | 55.56 | 81.77 | **82.09** |
| Bijectivity | 59.12 | 32.23 | 13.16 | 14.04 | **3.32** | 4.09 |
| Continuity | 1.499 | **1.418** | 1.531 | 1.496 | 1.896 | 1.735 |

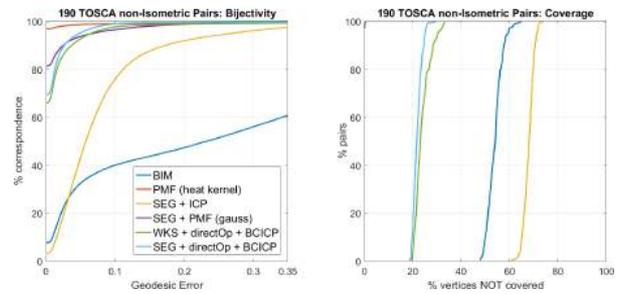

Fig. 17. Bijectivity (distance to identity) and coverage comparison on the TOSCA dataset: 190 pairs of non-isometric shapes.

Table 6. Runtime for meshes with $n$ vertices.

| Algorithm | Complexity | $n=1k$ | $n=5k$ | $n=10k$ | $n=15k$ |
|---|---|---|---|---|---|
| Bijectivity | $O(k^3)$ | 0.0663 | 1.7595 | 7.5121 | 18.831 |
| Continuity | $O(n)$ | 0.0339 | 0.2649 | 0.7120 | 1.2594 |
| Fix Outliers | $O(n)$ | 0.0073 | 0.0332 | 0.0851 | 0.1583 |
| Coverage | NA | 4.7305 | 5.6101 | 21.937 | 32.471 |

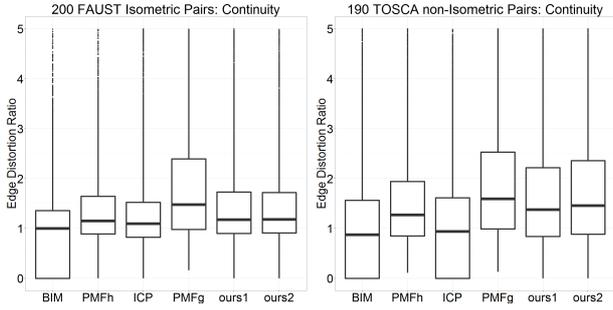

Fig. 18. Smoothness: 200 FAUST isometric pairs and 190 TOSCA non-isometric pairs. Here we show the edge distortion ratio of the following methods: BIM, PMF with heat kernel (PMFh), SEG + ICP (ICP), SEG + PMF with gauss kernel (PMFg), WKS + directOp + BCICP (ours1), SEG + directOp + BCICP (ours2). The difference between ICP and ours is small, but ours has a much higher coverage rate, without hurting the smoothness.

### 5.4 Parameters & Performance

*Runtime.* Let us first note that adding the orientation-preserving or reversing terms to the functional map pipeline does not change the runtime complexity, as it still leads to a least squares problem. The BCICP refinement pipeline includes four parts: (1) Bijective ICP: the time complexity is $O(k^3)$ for solving a linear system with $k$ basis functions and $O(n \log n)$ for nearest neighbor search where $n$ is the number of vertices. (2) Smoothing a map to improve the continuity: the time complexity is $O(cn)$, where $c$ is the maximum number of neighbors in the mesh. (3) Fixing outliers: we find the connected components which has complexity of $O(n)$ for meshes. (4) Coverage: the time complexity to improve the coverage depends on the quality of the input maps.

The runtime for meshes with different number of vertices is reported in Table 6, where for each category, we randomly picked 10 pairs of shapes from the SHREC dataset and measured the average runtime for each step (initialized by BIM). In all of our tests, we run the BCICP refinement for 5 iterations. The runtime of 8 random pairs of shapes (2 for each dataset) using other methods are reported in the supplementary materials.

*Convergence.* We do not have a theoretical analysis of the convergence rate of BCICP since some of the components are heuristic and are not designed to minimize this energy directly (e.g., removing the outlier region). However, our main observation is that these components combined together can help fix the major issues affecting the map quality, within a small number of iterations.

Fig. 19 shows an example of the convergence rate of BCICP (solid lines) compared to ICP (dashed lines) w.r.t. different measurements including the energy defined in Eq. (4) (orange), the fraction of uncovered vertices (dark blue), the bijectivity error (as defined in the "Measurement" paragraph above) (light blue), and the ratio of vertices that are categorized as outliers (as defined in Section 4.2.3) (green). More examples can be found the in the additional materials. Note that all the values in this plot have been normalized w.r.t. its corresponding initial values (specifically, the initial map has energy value 99.37, un-coverage 81%, bijection error 0.318, and the outlier ratio 17%). Note that although the energy curves are not monotonically decreasing, we can observe a major improvement in map quality (w.r.t. four different measurements) especially within the first 3 iterations, which is true for the vast majority of the tested pairs, and is not the case for ICP. The corresponding maps are visualized as inset of Fig. 19.

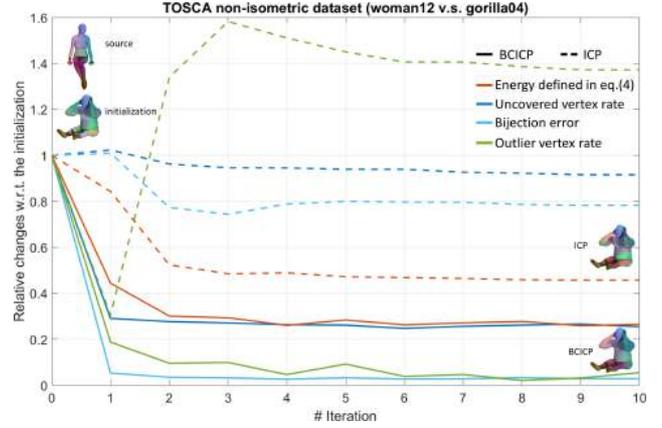

Fig. 19. Convergence. Here we show how the energy defined in Eq. (4) gets improved over 10 iterations using our BCICP refinement step (the solid orange line) and the regular ICP refinement (the dashed orange line). Some other important measurements are also included: the ratio of the vertices/areas that are not covered by the map (the dark blue lines), the bijection error (the light blue lines) and the ratio of the vertices/areas that are categorized as outliers (the green lines). All the measurements are normalized by the value at the initialization to make them visually comparable. Therefore, the plot shows the relative improvement over the initialization. The inset figures shows the initial map, and the resulting map of BCICP and ICP after 10 iterations.

*Parameters.* In our experiments, all the weights and parameters in the complete pipeline are set to the same values across all experiments. Specifically, for the functional maps pipeline (fMap) we used the default settings for the different regularizers provided by the authors. For the orientation-preserving/reversing term, the weight $\alpha_4$ is set such that this term has similar scale to the multiplicative term. We use 50 basis functions for both the source and the target shape to compute a functional map. For BCICP refinement, we use the following settings: the weights for the different energy terms in Eq. (4) are set to one. The number of basis that are used in the refinement step is 50 for the source and the target shape. The threshold for turning off the smoothness step is 0.6, i.e. if the coverage rate is below 0.6, we skip the smoothness step to avoid over-smoothed maps and a long running time. The threshold to evoke the coverage-promotion step is set to 0.5, i.e., if the coverage rate for the vertex-to-vertex map is below 0.5, we would improve the coverage to make the linear system more stable. The threshold in outliers detection is set as the maximum edge length, i.e., if the distance between the two mapped endpoints is larger than the maximum edge length on the target shape, this edge will be removed from the connectivity matrix.

Moreover, our framework takes the WKS descriptors or the segmentation as initializations. We used 50 eigenvectors to compute the WKS descriptors except for the TOSCA non-isometric dataset:

Table 7. Failure case. The table lists the percentage of pairs where our method (WKS/SEG + directOp + BCICP) produces worse results than the baseline methods: BIM or SEG + ICP.

| Dataset | compared to BIM (%) | | | compared to SEG + ICP (%) | | |
| --- | --- | --- | --- | --- | --- | --- |
| | perVtx | perMap | direct | perVtx | perMap | direct |
| Faust-Iso | 0.50 | 1.50 | 1.00 | 20 | 20 | 20 |
| Faust-nonIso | 1.25 | 8.00 | 7.00 | 7.75 | 5.75 | 5.75 |
| Tosca-Iso | 7.22 | 13.03 | 10.56 | 11.62 | 11.23 | 10.04 |
| Tosca-nonIso | 7.37 | 12.63 | 7.89 | 11.05 | 14.21 | 10.53 |

we used 250 eigenvectors instead given that this is the most challenging dataset and we need more information to compute the WKS descriptors. For segmentation computation, we used the default setting provided by the authors.

*Map coupling with projection.* As mentioned in Section 4.2.1, our BCICP procedure includes a coupling step (lines 13-15 of Algorithm 2) that updates the functional maps $C_{12}$ and $C_{21}$ using a projection of their composition. In practice, this step does not play a strong role when the intialized maps are already consistent, but can be particularly useful for non-isometric shapes and weak initializations.

In Figures 10–13, we plot the results of our method with (solid lines) and without (dashed lines) these coupling steps. Note that the difference between the solid and dashed lines is neglectable except for the TOSCA non-isometric dataset w.r.t the perMap and direct error measure. Nevertheless, we leave the exploration of other possible approaches for functional map coupling during refinement as an interesting direction for future work.

*Failure cases.* Table 7 shows the failure rate of our algorithm (WKS or SEG with direct operator and BCICP refinement). As failure we consider a shape pair for which our algorithm gives worse result than the baseline methods: regular ICP with SEG initialization and BIM. Specifically, for each of the datasets, we pick our winning method, i.e, SEG + directOp + BCICP for the FAUST dataset and WKS + directOp + BCICP for the TOSCA dataset. Then we measure the failure rates w.r.t. the two baseline methods. Moreover, for the TOSCA dataset, SEG + BCICP is worse than SEG + ICP for fewer than 3% of the shape pairs.

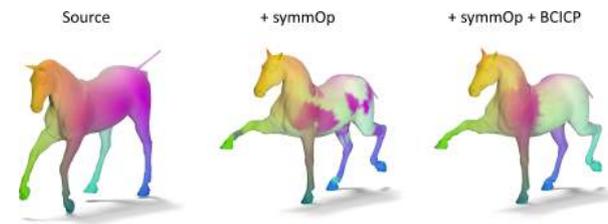

Fig. 20. Failure case. When the orientation-reversing term is not strong enough, the initial map (+symmOp) still contains a lot of symmetry ambiguities. Specifically, the front legs are roughly aligned with the direct correspondences, while the back legs are roughly aligned with the symmetric correspondences. In this case, the BCICP refinement can fail to produce a good map with consistent orientation.

### 5.5 Texture transfer & Post-processing evaluation

We also illustrate our results via texture transfer. In Fig. 21 we compare the results for various automatic methods on two isometric shape pairs from the TOSCA dataset. In Fig. 22 we show a comparison of deblur, ICP + CPD, and BCICP. In addition, in Fig. 23 we compare CPD and deblur to BCICP as a post-processing step where the initial maps are computed from functional map pipeline with 3 or 8 landmarks. These qualitative results illustrate the performance of our approach in selected challenging settings. They also highlight the fact that improvements in the chosen quantitative metrics result in visual improvements in practical applications.

## 6 LIMITATIONS AND FUTURE WORK

Our method has several limitations that we would like to overcome in future work. First, our method is very good at computing solutions that trade off different requirements: matching quality, bijectivity, coverage, and smoothness. However, we cannot enforce exact bijectivity or complete coverage. As our experiments with PMF show, enforcing strong bijectivity typically leads to many undesirable outliers. It would therefore be interesting to extend the current state-of-the-art to enforce bijectivity without generating outliers. Second, our approach is not directly applicable to point clouds or partial shapes. In future work, we would like to extend our work to a broader range of inputs so that we can directly process point clouds stemming from scanned data. Third, shape matching on shape collections typically works for smaller meshes with up to 10k vertices. It would be rewarding to derive very fast shape matching algorithms that can tackle shapes that are orders of magnitude larger in the future, by, e.g., parallelizing our map refinement steps.

## 7 CONCLUSIONS

In this paper, we proposed an algorithm to compute orientation-preserving maps via the functional maps pipeline. We first introduced a novel term for computing functional maps that promotes orientation preservation directly in the functional (spectral) domains. We then extended the iterative post-processing pipeline to improve maps both in the spectral and spatial domains. We demonstrated the advantages of our method by comparing to several recent state-of-the-art methods on well known test datasets. Our framework not only improves the quality of the maps compared to ground truth, but also the bijectivity and continuity of the maps.

### ACKNOWLEDGMENTS

The authors would like to thank the anonymous reviewers for their valuable comments and helpful suggestions. The authors would like to thank Zorah Lähner, Danielle Ezuz, Emanuele Rodolà, Dorian Nogneng, Dongming Yan, and Ruqi Huang for providing code and valuable discussions. This work was supported by the KAUST Office of Sponsored Research (OSR) under Award No. OSR-CRG2017-3426, the Jean Marjoulet chair from Ecole Polytechnique, a Google Research Award, and the ERC Starting Grant No. 758800 (EXPROTEA).

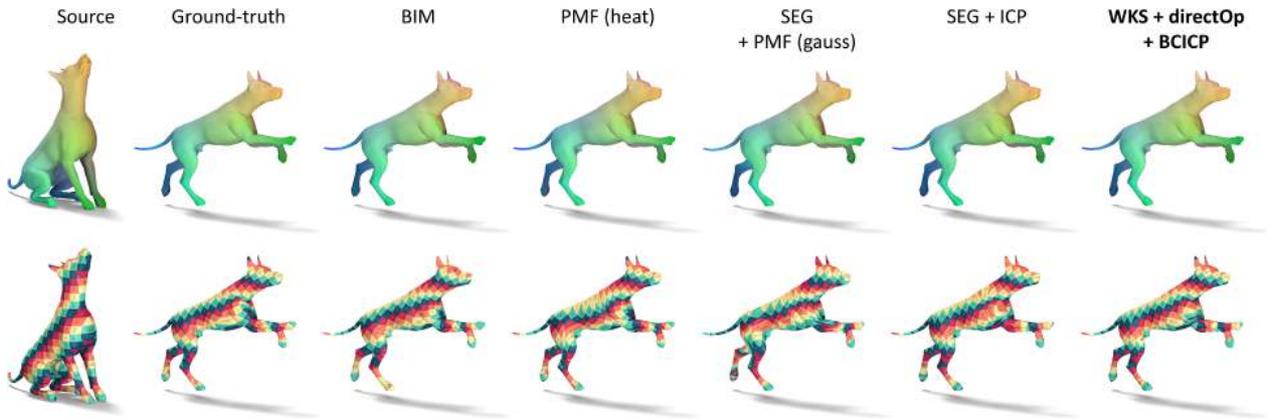

Fig. 21. Texture transfer: We compare our algorithm to different competing methods. We use the vertex-to-vertex maps (visualized in the first row) to transfer the texture to the target shape (visualized in the second row). The texture-transfer provides a good visualization for the local distortion of the resulting maps.

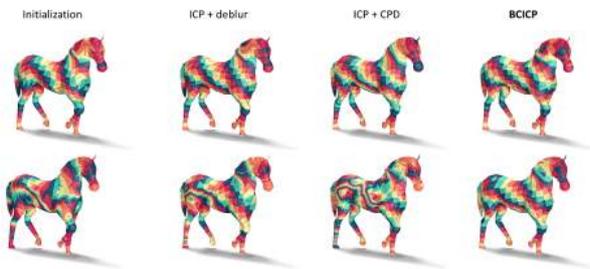

Fig. 22. Texture transfer: visual comparison of deblur, CPD, and BCICP as a post-processing step. The first column shows two initializations with different quality. Then we compare our BCICP refinement with two other methods. CPD and deblur can work well when the initialization has high quality, but they may fail when the initialization is not good. Our BCICP refinement is more stable w.r.t. initializations.

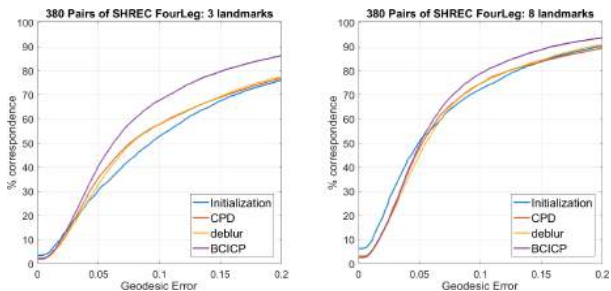

Fig. 23. Post processing: compare the CPD, deblur and BCICP as a post-processing step, where the initial maps are initialized with 3 or 8 landmarks.

## 8 APPENDIX

Proof of Theorem 4.1.

PROOF. Given a compact oriented surface $M$ without boundary, our goal is to show that for any smooth $f : M \to \mathbb{R}$ and $g : M \to \mathbb{R}$, we have:

$$\int_{p \in M} \langle \nabla f(p) \times \nabla g(p), \mathbf{n}(p) \rangle d\mu(p) = 0, \quad (11)$$

where $\mathbf{n}(p)$ is an outward pointing normal at a point $p$, while $\times$ is the standard vector cross product in $\mathbb{R}^3$. Indeed, if we consider the volume $V$ bounded by $M$, and functions $\tilde{f}, \tilde{g} : V \to \mathbb{R}$ whose gradients agree with those of $f$ and $g$ on $M$, then we have, by the divergence theorem:

$$\int_{p \in M} \langle \nabla f(p) \times \nabla g(p), \mathbf{n}(p) \rangle d\mu(p) = \int_{q \in V} \mathrm{div}\left(\nabla \tilde{f}(q) \times \nabla \tilde{g}(q)\right) d\mu(q)$$

$$= \int_{q \in V} \left( \langle \mathrm{curl}(\nabla \tilde{f}(q)), \nabla \tilde{g}(q) \rangle - \langle \mathrm{curl}(\nabla \tilde{g}(q)), \nabla \tilde{f}(q) \rangle \right) d\mu(q) = 0.$$

Here the second to last equality holds by the basic properties of vector products, while the last one follows from the curl-free nature of gradients. □

Interestingly, the same statement also holds in the discrete setting, when normals are given on faces of the triangle mesh and the gradients are discretized by linear interpolation of function values at the vertices. A basic calculation in that case shows that for any closed manifold mesh, we have: $\sum_{t \in \text{triangles}} \langle \nabla f(p) \times \nabla g(p), \mathbf{n}(p) \rangle A(t) = 0$.